\documentclass[12pt]{iopart}
\usepackage{graphicx}
\usepackage{hyperref}

\begin{document}

\note{Estimation of phase in EEG rhythms for real-time applications}









\author{J. R. McIntosh\textsuperscript{1}, P. Sajda\textsuperscript{1, 2}}

\address{\textsuperscript{1}Department of Biomedical Engineering, Columbia University, New York, NY 10027}
\address{\textsuperscript{2}Data Science Institute, Columbia University, New York, NY 10027}
\ead{j.mcintoshr@gmail.com}
\vspace{10pt}
\begin{indented}
\item[]October 2019
\end{indented}

\begin{abstract}
\textbf{\textit{Objective.} We identify two linked problems related to estimating the phase of the alpha rhythm when the signal after a specific event is unknown (real-time case), or corrupted (offline analysis). We propose methods to estimate the phase prior to such events. \textit{Approach.} Machine learning methods are used to mimic a non-causal signal-processing chain with a purely causal one. \textit{Main results.} We demonstrate the ability of these methods to estimate instantaneous phase from an electroencephalography signal subjected to very minor pre-processing with higher accuracy than more standard signal-processing methods. \textit{Significance.} Phase estimation of EEG-rhythms is a challenge due to non-stationarity and low signal to noise ratio. The methods presented enable scientists and engineers to achieve relatively low error by optimizing causal phase estimation on a non-causally processed signal for a real-time experiments and offline analysis.}
\end{abstract}

\vspace{2pc}
\noindent{\it Keywords}: Electroencephalography (EEG), Real-time, Closed-loop, Phase estimation, Alpha rhythm, Causal filtering


\section{Introduction}

Techniques that make use of real-time estimates of the phase of endogenously generated brain rhythms to trigger transcranial magnetic stimulation (TMS) pulse delivery \cite{Bergmann2012_jns,Chen2013_tbme,Safeldt2017-mi,Faller2019-vj} are of increasing interest, and starting to be used in neuroscience and clinical studies \cite{Bergmann2012_jns,Faller2019-vj,Zrenner2016-xt,Thut2017_clinneuro,Zrenner2018-fa,Bergmann2018,Madsen2019-cd,Desideri2019-ya}. Here we investigate a machine learning approach to potentially improve phase estimation of brain rhythms, where low signal-to-noise (SNR) and non-stationarity are significant, and causality must be respected.

In the main scenario considered in this manuscript (the real-time case) we are interested in knowing the current instantaneous phase so that we can decide whether or not to generate an event, such as firing a TMS pulse, or presenting a stimulus. In a related scenario, that event has already happened (the offline analysis case), and we are interested in knowing the estimate of phase at that moment. This may be of interest, for example in order to explain brain activity and behavior after an event, given a rhythm’s phase at the moment of that event \cite{Keil2014-tl,Van_Elswijk2010-we,Busch2009-yz,Mathewson2009-di,Thut2011-jj,Ng2012-ad,Scheeringa2011-gq,Dugue2011-by}. However, the estimate of phase in this scenario has to be made with care to avoid systematic errors caused by the events themselves, or their influence on brain activity \cite{Zoefel2013-mf,Widmann2015-to,Vanrullen2011-ev,Acunzo2012-pi,Tanner2015-xk}. These two scenarios are identical from a signal processing point of view as they must both be treated with methods that do not make use of information after the event of interest.

\begin{figure}
  \includegraphics[width=1.0\textwidth]{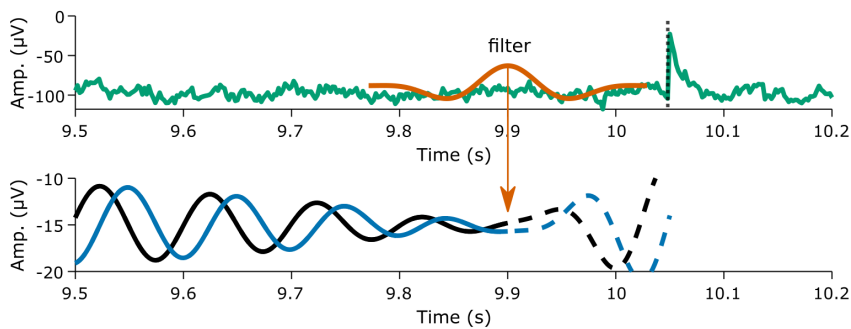}
  \caption{Application of an FIR filter on an EEG signal (green) must not include the event itself (dashed line), but consequently results in a group delay since the estimate of phase  (represented as $sin(\phi)$ in blue and $cos(\phi)$ in black) lags the leading filter edge. We propose learning the relationship between the raw data and the instantaneous  phase in data stretches that do not include events so that they may then be applied to predict the phase prior to such events.}
  \label{fig_intro}

\end{figure}

An example of a standard approach to recover phase from a signal \cite{Keil2014-tl,Thut2011-jj,Zoefel2013-mf} is to use a finite impulse response (FIR) bandpass filter (causally applied), followed by a Hilbert transform. Given that the filter window should not overlap the event of interest (either because there is no signal, or because the event would corrupt the phase estimate, see Fig.  \ref{fig_intro}), the estimate of phase comes at a time point prior to the event which is related to the order of the filter. To extract the phase at the event time, one might choose to linearly extend the phase forward in time with a slope provided by an estimate of the oscillation frequency. Four related problems arise with such a method 1) frequency, amplitude and waveform shape may not be stationary properties of a specific brain rhythm \cite{Soernmo2005,Cole2019-ok,cole2017brain} 2) estimates of the instantaneous frequency are noisy 3) FIR filter order should be long to adequately extract the band of interest 4) due to 1), and in contrast to 3) FIR filter order should be short to minimize reliance on the (usually simple) model of phase propagation. Similar methods that make use of wavelet-based and Fourier signal analysis have also been used, but they are unlikely to differ substantially from the example presented above \cite{Bruns2004-pc}.

To address the stated problems, we attempt to recover EEG alpha rhythm phase during an auditory oddball paradigm \cite{Hong2014-wo} from the POz electrode prior to a specific event. Specifically, we begin by learning the relationship between the minimally pre-processed EEG signal, and the non-causally processed instantaneous phase. This learned relationship is then used to predict the phase at the instant just prior to an event. We show that this method provides considerably better phase estimates than a non-machine learning based approach. We go on to examine the recovery of ground-truth in simulated data. More generally, in this manuscript we argue in favor of applying machine learning approaches to map the relationship between raw signals and non-causally recovered states in a causal manner, for applications in real-time operation, as well as general offline analysis.

\section{Methods}
\subsection{Data description}
The data used for this study was first published in \cite{Hong2014-wo},to which we refer the reader for further details and motivation. Fifteen subjects took part in the experiment. One subject was excluded due to abnormalities in the data. An auditory oddball paradigm (80\% standard, 20\% target) was used. Stimuli were presented through speakers and each lasted for 200ms with an inter-trial interval (ITI) sampled from a uniform distribution between 2s and 3s. Subjects were instructed to press a button on a gamepad with their right index finger as soon as they heard the target sound. Subjects took part in 5 runs, each consisting of 75 target trials and 300 standard stimulus presentations. During the task, EEG was recorded at 2048Hz using a 64 scalp electrode ActiveTwo system (Biosemi, The Netherlands) with electrodes in the extended 10-20 configuration. This study was carried out in accordance with the guidelines and approval of the Columbia University Institutional Review Board. Written informed consent was obtained from all participants.

\subsection{Pre-processing}
The MNE package \cite{Gramfort2013-la,Gramfort2014-yr} was used to load EEG recordings into Python (v3.7). We non-causally low-pass filtered all EEG data at 50Hz to remove any remnant of line noise, and resampled to 512Hz. These steps are likely to be unnecessary for the application of our method, as would be the preferred usage in a real-time setting, although substantial line noise could encourage learnt filters to favor noise suppression. In order to extract a local oscillation as might be done in a real-time setting \cite{Zrenner2018-fa} we applied a Hjorth \cite{Hjorth1975-ig} centred at POz which entailed subtracting the average of the signal at PO3, PO4, Pz and Oz from POz. We estimated the individual alpha frequency (IAF) of each subject by taking the frequency at the maximum power in the 8-12Hz range. In order to aid generalization of our method across subjects, data was standardised by dividing by the average power in the 30-50Hz band to make our noise floor consistent across subjects.
\begin{figure}
  \includegraphics[width=1.0\textwidth]{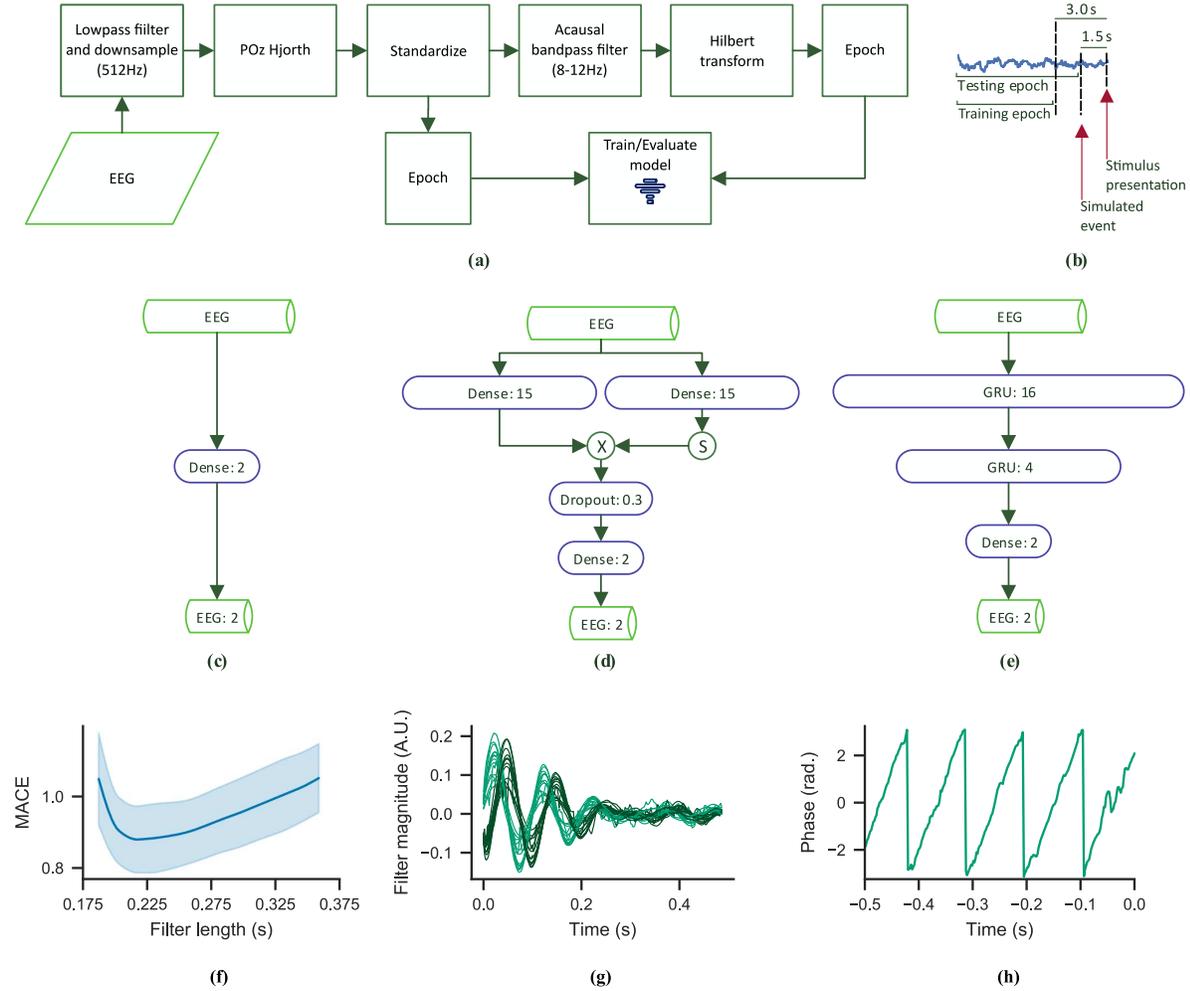}
  \caption{Processing pipeline and proposed architectures. \textbf{(a)} Processing pipeline as described in methods section \ref{sec_traintest}. \textbf{(b)} Data splitting for training vs testing. The simulated event at which we evaluate phase occurs 1.5s before the real event (target stimulus presentation) in order to avoid a systematic disturbance. For training, the epoch is cut a further 1.5s prior. \textbf{(c)} Optimised filter (OF) \textbf{(d)} Optimized multi-layer filter architecture (MLOF) \textbf{(e)} Gated recurrent unit (GRU) architecture. \textbf{(f)} Mean absolute circular error (MACE) for different FIR filter lengths. Lowest MACE is for a filter of length 0.22s, which presumably offers a balance between noise suppression and distance to event. Shaded regions represents $2\times$ SEM calculated across subjects. \textbf{(g)} Example weights for the optimized filter method, individual traces represent a single subject, while colors represent different filters. \textbf{(h)} Example phase estimates prior to an event (t = 0) for the optimized filter method.}
  \label{fig_methods}
\end{figure}

\subsection{Model training and testing}\label{sec_traintest}
After pre-processing, the signal processing pipeline splits along two paths (see Fig.  \ref{fig_methods}a). On one branch data was epoched to generate the input signal (the features used for training, or the features on which we evaluate phase estimation). On the other branch, we applied an FIR bandpass filter (8-12Hz passband, order 769, approximately 1.5s) and took the Hilbert transform prior to epoching in order to generate the output signal (used as the output for training, or as a signal against which we evaluate our phase estimation).

Epoching proceeded as shown in Fig. \ref{fig_methods}b. During training we used entire data stretches between events up to 1.5s prior to each simulated event. At test time, we evaluated phase on the same stretches of data, but extended the prediction up to the simulated event itself.

We based our simulated event on the target stimulus presentation. However, because we are interested in evaluation rather than actually recovering phase at these events, in this work we made the following two modifications: 1) Simulated events were shifted to be 1.5s prior to presented target events. This is so that after non-causally filtering, the event itself could not interfere systematically with the phase prediction. 2) We only included every other stimulus presentation in order to allow sufficient data in each trial per run. It should be noted that for applications of our method rather than evaluation, neither of these steps would be taken. The way in which these epochs are used then depends on the method under consideration as discussed in the following sections.

\subsection{Compensated FIR filter methods}
In order to remove high amplitude drift which is difficult for a short filter to suppress, we initially processed the signal by causally subtracting a 2s moving average. The input signal epochs were then causally bandpass filtered up to the event time, and the phase was estimated with a Hilbert transform. The bandpass filter for our standard method (C-FIR, 0.22s) takes the form of a FIR filter of 0.22s length with an 8-12Hz passband applied causally. The filter length of 0.22s was set as a compromise between a long filter necessary to suppress noise, and a short filter which is more able to deal with a non-stationary signal (due to its lower group delay, see Fig. \ref{fig_methods}f). We also introduced an IAF based filter (IAF C-FIR, 0.22s) with a 3Hz passband centred around the IAF of the subject. After the Hilbert transform is applied to the signal, phase was extracted five samples prior to the end of the epoch to reduce edge effects. The IAF of the subject was then used to compensate for the bandpass filtered group delay and data removed to cater for edge effects. No training step is required for these methods, so they were directly evaluated on the data epoched for testing.

\subsection{Learning based prediction}
Unlike in the FIR based approach, we suppressed electrode drift by calculating the backwards difference \cite{oppenheim2010} and using that as our signal. We found that this was an effective method to suppress low frequencies that sped up learning, although it was not strictly necessary. For these methods, the architectures were trained with an Adam optimizer \cite{Kingma2014-vo} to minimise the mean squared error (MSE). Training was done on an NVIDIA GeForce GTX 1080 GPU with  CUDA  10.0  and  cuDNN  v7.1,  in  Tensorflow v1.14 \cite{Abadi2016-cu}, using the Keras API \cite{Chollet2015-ca}, and CuDNNGRU layers were used to speed up training. 20\% of the training set was held out and treated as validation data which is used to halt training if MSE did not reduce for 12 consecutive training epochs, with the model matching the weights generating the lowest MSE on the validation set being finally selected.

Fig. \ref{fig_methods}c-e describe in full the model architectures we investigated for learning the relation between the input signal and the output signal. Specifically, Fig. \ref{fig_methods}c corresponds to our standard optimized filter (OF). For this method weights of a single dense layer are learnt to map from the raw data, to the cosine and sine of the instantaneous phase, recovered via non-causal filtering and application of the Hilbert transform. Because of the use of a single dense layer, this is similar to performing multivariate multiple linear regression on the Toeplitz matrix of the input signal, while using the same output signal. The example weights that are learnt by training the model on all data from each subject are shown in Fig. \ref{fig_methods}g, with light green corresponding to $sin(\phi)$ and dark green corresponding to $cos(\phi)$, and Fig. \ref{fig_methods}h shows an example of the phase progression as estimated by this method. Fig. \ref{fig_methods}d corresponds to the optimized multi-layer filter (MLOF). We considered that the OF should not, in general, be optimal at capturing phase when the underlying frequency of the signal of interest in unstable. We consequently considered neural-network based methods with non-linearities which could capitalise on the frequency information. This network was constructed so that it would be capable of learning to suppress some filter outputs and enhance others dependent on the mix of frequencies present by the introduction of multiple dense layers capable of gating each other via softmax operations. Finally, Fig. \ref{fig_methods}e corresponds to an approach that makes use of gated recurrent units (GRU) \cite{Chung2014-zv}. Motivated in the same way as the MLOF method, and because instantaneous phase estimation can be viewed as a state estimation problem where observations of the alpha rhythm should be integrated optimally with predictions of the future phase, we attempted to use GRUs, which are a type of recurrent neural network (RNN) \cite{Goodfellow2016-ut,Rumelhart1988-ds}.

\subsection{Simulated data}
With recorded data we can only compare our causal estimate of phase to a non-causal estimate of phase. In reality the non-causal phase estimate may still be a poor estimate of phase due to the presence of autocorrelated noise, and it would consequently be useful to calculate how our different methods perform against a ground-truth signal. A ground-truth signal is simply not available for recorded EEG data, so we resort to modeling the EEG and alpha-rhythm to investigate how well the ground-truth phase of the underlying model can be recovered.

We chose to model the EEG signal with a Kuramoto model \cite{Kuramoto1984}, in which the phase of an individual oscillator progresses dependent on its own natural frequency and its interactions with the natural frequency of the population. We chose this as it can demonstrate dynamics reminiscent of the alpha rhythm, such as changing frequencies, and varying amplitudes \cite{Breakspear2010-ty}. Ground-truth phase was defined as the non-causally recovered phase after applying a Hilbert-transform, prior to the addition of noise.

In order to constrain the Kuramoto model fits, we draw N = 16 driving frequencies (corresponding to oscillators) from a Cauchy distribution centred on the subject’s IAF ($f_{IAF}$): $w_i \sim Cauchy(2 \pi f_{IAF}, \gamma)$. The phase of the oscillators develop according to the following equations:

$$\Delta \theta_i = \omega_i + \frac{K}{N}\sum_{j}^{N}(\theta_j (t) - \theta_i (t))$$
$$\theta_i(t+\Delta t) = \theta_i(t) + \Delta\theta_i + c\xi \sqrt{(\Delta t)}$$

We then chose to define the the full simulated alpha rhythm $s(t)$ as:
$$s(t) = A\sum_i^N cos(\theta_i (t))$$

Where $\theta(t)$ represents the instantaneous phase of each oscillator, initial spread $\gamma$, amplitude $A$, coupling $K$ and noise scaling term $c$ on a Gaussian process $\xi \sim N(0, 1)$.
The full simulated EEG signal then takes the form $y(t) = s(t) + p(t)$, where $p(t)$ represents a signal generated from a $1/f^k$ noise generator fitted to match the power spectrum of the individual subject using the FOOOF toolbox \cite{Haller2018-mj}. Parameters $\gamma, A, K, c$ are then fitted to generate a (power spectral density) PSD that match the PSDs of individual subjects: Optimisation was implemented by drawing from uniform distributions of each of the parameters to be fitted. We evaluated the PSD for the signal generated from the Kuramoto model with randomised parameters 500 times. We then picked the parameter set with the lowest cost, where cost is defined as the absolute difference in the power spectrum between 2.5 and 30Hz. The parameter distributions that were drawn from are defined as follows: $\gamma \sim [0, 2\pi]$, $A \sim [0.15, 100]$, $K \sim [0, 20]$, $c \sim [0, 2\pi]$.

\subsection{Statistics}
Mean absolute circular error (MACE) between ground-truth or non-causally predicted phase phase and causally predicted phase was calculated across events for each method. In order to compare MACE between methods, we employed a Wilcoxon signed-rank test and used Bonferroni correction for multiple comparisons unless stated otherwise. The median difference (MD) between MACE values was also calculated.


\section{Results}
\subsection{Prediction on recorded data}
Our premise is that the instantaneous phase of neural oscillations can be estimated with higher accuracy by using methods rooted in machine learning. Specifically, the causal relationship between an EEG signal and a non-causally derived target can be learnt and then applied to regions of the signal that were otherwise inaccessible. We compared three direct learning based approaches to two standard methods that may be used to recover phase.

\begin{figure}
  \includegraphics[width=1.0\textwidth]{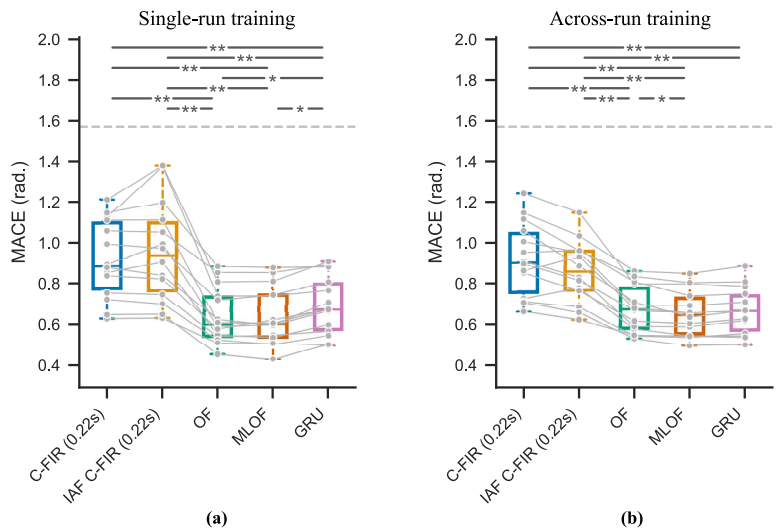}
  \caption{Mean absolute circular error (MACE) is lower for all learning methods compared to C-FIR methods. \textbf{(a)} Training on one run, and evaluating on the subsequent run. Comparison of MACE for a generic alpha-band 0.22s compensated FIR (C-FIR) filter (blue), IAF tuned 0.22s C-FIR filter (yellow), optimized filter (OF, green), optimized multi-layer filter (MLOF, orange), GRU (purple). Connecting lines (gray) represent subjects. Hinges represent first and third quartile and whiskers span the range of the data not considered outliers. Asterisks between box-whisker plots represent statistical significance level after Bonferroni corrected Wilcoxon signed-rank tests (see Table \ref{table_wr} for statistics). \textbf{(b)} Training is performed on all subjects apart from one held-out subject and following evaluation is performed on the held-out subject. Measures as in (a), see Table \ref{table_as} for statistics.}
  \label{fig_res_data}

\end{figure}

We found that the learning based approaches consistently showed mean absolute circular error (MACE) values which were lower than compensated FIR methods. This was true regardless of whether training was performed on a single run and evaluated on the subsequent run of an individual subject ($p = 9.8 \times 10^{-4}$, averaged FIR methods, compared with averaged learning methods, Wilcoxon signed-rank test Fig. \ref{fig_res_data}a, Table \ref{table_wr}), or whether training was performed on all but one held-out subject, and then evaluated on that held-out subject ($p = 9.8 \times 10^{-4}$, averaged FIR methods, compared with averaged learning methods, Fig. \ref{fig_res_data}b, Table \ref{table_as}).

Between the direct learning methods, it seems that contrary to our predictions, the GRU based method did not achieve the lowest MACE (across-run training, $\mathrm{MD_{MLOF-GRU}} = -0.013$,  $p = 0.110$; $\mathrm{MD_{OF-GRU}} = 0.010$, $p = 1.00$). On the other hand, for the across subject condition the MLOF outperformed the OF method as expected (across-run training, $\mathrm{MD_{OF-MLOF}} = 0.023$, $p = 0.015$), potentially due to its ability to perform a type of frequency selection. This improvement however is rather small in comparison to the improvement of any of the learning methods over the more standard methods.

\begin{table}
\caption{\label{table_wr}Median differences (MD) and Bonferroni corrected statistical results for within-subject analysis, presented in the form MD/$p$.}
\begin{indented}
\item[]\begin{tabular}{@{}lllll}
\br
MD/\textit{p}   & IAF C-FIR & OF    & MLOF  & GRU   \\
\mr
C-FIR       & -0.037/1.00   & 0.277/0.010   & 0.278/0.010   & 0.227/0.010  \\
IAF C-FIR   &               & 0.314/0.010   & 0.314/0.010   & 0.264/0.010  \\
OF          &               &               & 0.001/1.00    & -0.050/0.019  \\
MLOF        &               &               &               & -0.051/0.015  \\
\br
\end{tabular}
\end{indented}
\end{table}

\begin{table}
\caption{\label{table_as}Median differences (MD) and Bonferroni corrected statistical results for across-subject analysis, presented in the form MD/$p$.}
\begin{indented}
\item[]\begin{tabular}{@{}llllll}
\br
MD/\textit{p}   & IAF C-FIR & OF    & MLOF  & GRU   \\
\mr
C-FIR       & 0.058/0.132   & 0.242/0.010  & 0.264/0.010  & 0.252/0.010  \\
IAF C-FIR   &               & 0.183/0.010  & 0.206/0.010  & 0.193/0.010  \\
OF          &               &               & 0.023/0.015  & 0.010/1.00  \\
MLOF        &               &               &               & -0.013/0.110  \\
\br
\end{tabular}
\end{indented}
\end{table}

\subsection{Prediction on model data}
\begin{figure}
  \includegraphics[width=1.0\textwidth]{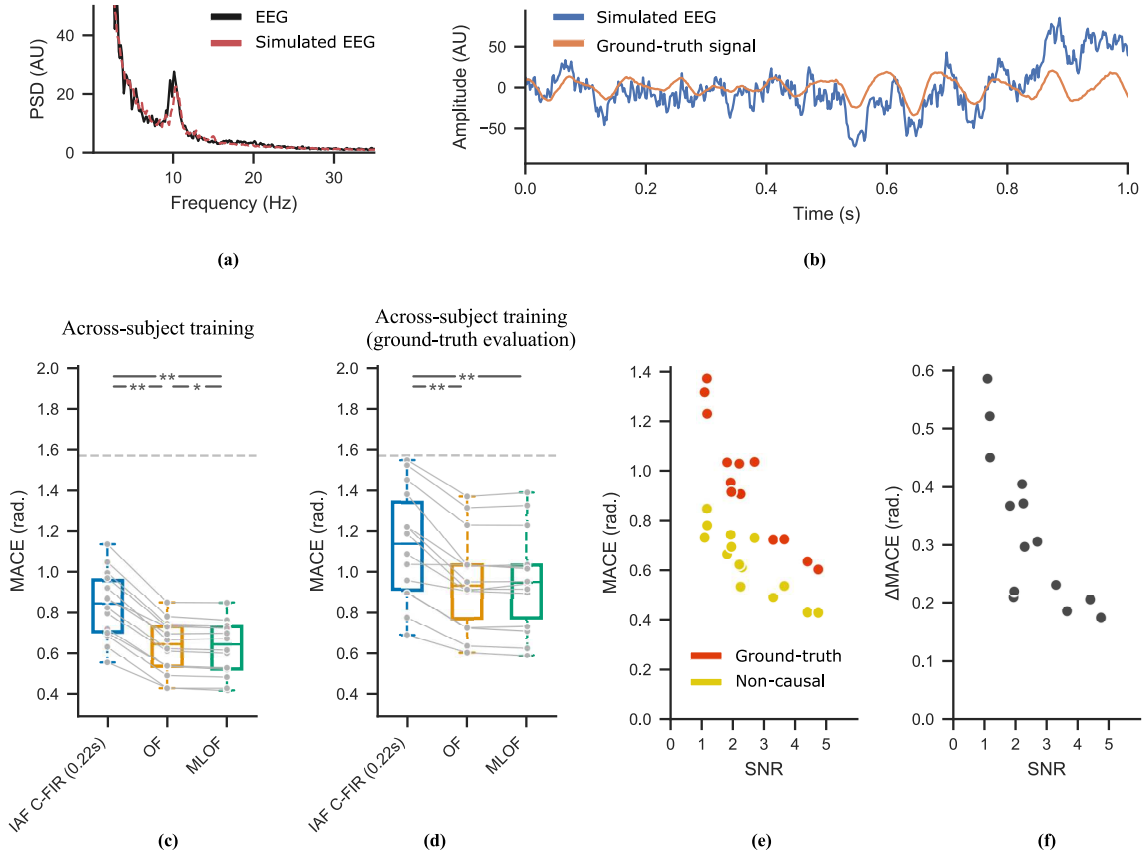}
  \caption{Phase predictions on data simulated to generate power spectra that match real data. As for real data, all direct learning methods outperform C-FIR methods. \textbf{(a)} Example PSD plot for a single subject (black), with PSD of matching simulated data (red). \textbf{(b)} Example simulated time series (blue) with underlying ground-truth ‘alpha’ oscillation overlayed (orange). \textbf{(c)}  Held-out simulated subject training, equivalent to Fig. \ref{fig_res_data}b. \textbf{(d)} In simulated data, unlike in recorded data the phase recovery performance can be compared to ground-truth. Phase is in general recovered relatively poorly with respect to ground-truth, although the direct learning methods still outperform C-FIR methods. Hinges represent first and third quartile and whiskers span the range of the data not considered outliers.Asterisks between box-whisker plots represent statistical significance level after Bonferroni corrected Wilcoxon signed-rank tests. \textbf{(e)} As SNR increases recovered MACE decreases when evaluation is done against non-causally recovered phase (as in (c), yellow) as well as against ground-truth recovered phase (as in (d), red). Shown for the OF method. \textbf{(f)} Difference between non-causally recovered phase and ground-truth recovered phase evaluation shown in (c). As expected, when SNR increases non-causal based evaluation approaches ground-truth evaluation.}
  \label{fig_res_simu}
\end{figure}

A valid objection to the results presented in the previous section is that the non-causal phase estimate is itself an estimate of the ground-truth phase. In this section, we attempt to address this by modeling the EEG with a Kuramoto model matched to each subject (Fig. \ref{fig_res_simu}a), where the underlying dominant phase can be estimated in the absence of noise (Fig. \ref{fig_res_simu}b) in order to provide a ground-truth measure. Armed with EEG generated from these Kuramoto models, we can investigate how well the ground-truth phase is recovered by the proposed methods. We did not evaluate the GRU method, or C-FIR (0.22s) further for clarity.

The general ranking of the direct learning models compared to the standard methods appears consistent with real data, with the exception that the OF produces a lower MACE than the MLOF method (across-run training with non-causal evaluation, $\mathrm{MD_{FIR-OF}} = 0.204$,  $p = 0.003$; $\mathrm{MD_{FIR-MLOF}} = 0.210$, $p = 0.003$; $\mathrm{MD_{OF-MLOF}} = 0.006$, $p = 0.040$, Fig. \ref{fig_res_simu}c, cf. Fig. \ref{fig_res_data}b). Models that were trained on phase recovered by non-causal processing of simulated EEG signals and evaluated against the ground-truth phase (Fig. \ref{fig_res_simu}d) have similarly low MACE for the learning methods (across-run training with ground-truth evaluation, $\mathrm{MD_{FIR-OF}} = 0.178$,  $p = 0.003$; $\mathrm{MD_{FIR-MLOF}} = 0.178$, $p = 0.003$; $\mathrm{MD_{OF-MLOF}} = -0.000$, $p = 1.00$). However what is generally apparent if our simulated data is considered a valid proxy for real EEG, is that all methods are in reality substantially worse than might be expected based on non-causal phase estimates when assessing their MACE ($p = 0.001$ for each method independently, not Bonferroni corrected, when ground-truth evaluation is compared to non-causal evaluation). To investigate this further, we examined how the SNR of the simulated EEG relates to MACE as shown in Fig. \ref{fig_res_simu}e. As expected, when SNR decreases, the MACE increases, with the ground-truth based MACE being higher than the non-causally filtered MACE. Interestingly, for low SNR the MACE is relatively more inflated (Fig. \ref{fig_res_simu}f) by the non-causal filtering method. This is consistent with the idea that when SNR is low, the phase of bandpass filtered noise is being recovered, while when SNR is high, the phase of the signal of interest is being recovered.

\section{Discussion}
In this manuscript we propose that for many practical implementations of instantaneous phase recovery, we can improve on standard signal-processing approaches by learning the mapping from the raw signal, to the instantaneous phase. Our primary aim has been to highlight the improvements available in real-time EEG applications that rely on phase estimates by deploying machine learning approaches.

\textit{Non-linear methods - } We were expecting that when evaluating our learned models across-subjects the GRU based approach would outperform all others, hypothesising that the model may be able to cater to the changing dominant frequencies both between and across subjects more readily. However, this was not the case. It may be that the GRU architecture requires features for state progression that are not readily accessible in shallow networks, or simply that the number of parameters to be learnt puts this approach at a disadvantage. On the other hand, we found small gains when applying the MLOF over the OF method in the across subject condition as expected, and we ascribe this improvement to the ability of the network to cater to the IAF of individual subjects by suppressing irrelevant frequencies. Having said this, the improvement conferred by this architecture is marginal when compared to the general improvement provided by employing learning methods. Because of this, it may be more practical to use the OF method, which can also be cast as a simple linear regression making it easy to implement in a variety of different scenarios.

\textit{Real-time operation and offline analysis - } Unlike other methods \cite{Safeldt2017-mi,Chen2013_tbme,Faller2019-vj}, our  approach has the disadvantage that some preliminary data must be collected prior to the main experiment. We found however, that at least for the alpha-rhythm, learning methods generalised well across subjects. This suggests that previous studies may be used to replace the initial data collection step on individual subjects, although we suggest that held-out, across subject validation as presented here should be performed, especially when attempting to generalise the presented results to other tasks or frequency bands.

Aside from the reduction in MACE provided by employing a machine learning approach which directly optimises on phase estimation, it could be argued that our method also introduces simplifications in the real-time implementation. For example, if there is a known system delay of 20ms between a collected sample, and the stimulation event time that has made use of that collected sample, then this delay can be built directly into the learned relationship between the raw signal and stimulation. This means that the phase estimation for the expected stimulation time can be collapsed to a single evaluation of the learning method on the raw data.

The application of directly learning the relation between the raw signal and instantaneous phase has uses beyond real-time experiments. In the real-time estimation of phase, we simply do not have future information, however in other situations where we are interested in the phase at specific event times (e.g. a TMS pulse, or a stimulus), while the signal is present past this event, it is corrupted by the event itself. In such a situation then, in order to not systematically corrupt the phase estimate dependent on the stimulus event itself we must use only the information present up to the event time. The method presented here is such an approach. As a caveat, the learned filters are directly dependent on the underlying properties and quality of the data and consequently do not guarantee unbiased phase estimates. Caution must therefore be used when analysing the distribution of generated phases, and this property must be traded off against the comparatively low MACE that they generate. Critically however, unlike in the case where phase is being recovered from application of non-causal filters the phase distribution is independent of the post-event activity.

\textit{Ground-truth recovery - } We were somewhat surprised at the relatively high MACE calculated based on ground-truth phase, compared to when it is calculated on non-causally filtered data. Our simulations match expectation of real data in showing that when SNR is low, phase estimation error is high, and show that non-causal filter based evaluation of phase is biased to appear better than it really is. More interestingly, they clearly show that this bias is relatively stronger at low SNR. In other words, estimates of phase evaluation error are artificially more inflated at low SNR, which may be of general practical importance.

\textit{Conclusion - } The principle laid out in this manuscript is to mimic a non-causal signal-processing chain with a purely causal one by using machine learning approaches. This enables scientists and engineers to optimise phase targeting in a real-time environment. The method may also be deployed in offline analysis as an alternative to non-causal filtering when estimation of phase is key, but runs the risk of being biased by an event starting at the time for which phase must be estimated. Finally, in future work this approach could be applied to predict features of EEG-rhythms \cite{Cole2019-ok} other than phase.

\ack
We thank Linbi Hong for data collection, and Josef Faller and Anna Jasper for comments. Funding: The National Institute of Mental Health under grant R33MH106775 and the United States Army Research Laboratory under Cooperative Agreement W911NF-10-2-0022. The views and conclusions contained in this document are those of the authors and should not be interpreted as representing the official policies, either expressed or implied, of the United States government.

\section*{References}
\bibliographystyle{unsrt}
\bibliography{references}
\end{document}